\documentclass[preprint]{revtex4-2}
\usepackage[dvipdfmx]{graphicx}
\usepackage{amsmath,amssymb}
\usepackage{txfonts}
\usepackage{bm}
\usepackage{ascmac}
\usepackage{bigints}
\usepackage{setspace}
\usepackage{mathptmx}
\usepackage[version=3]{mhchem}
\usepackage{color}

\usepackage{here}

\begin{document}
\title{Reverse non-equilibrium molecular dynamics simulations of a melt of Kremer-Grest type model under fast shear}
\author{Tatsuma Oishi$^{1,3}$,Yusuke Koide$^1$, Takato Ishida$^{1,2}$,Takashi Uneyama$^1$,Yuichi Masubuchi$^1$,Florian M\"{u}ller-Plathe$^3$}
\affiliation{\textsuperscript{1} Department of Materials Physics, Graduate School of Engineering, Nagoya University,
Furo-cho, Chikusa, Nagoya 464-8603, Japan}
\affiliation{\textsuperscript{2} Institute for Advanced Research, Nagoya University, Furo-cho, Chikusa, Nagoya 464-8601, Japan}
\affiliation{\textsuperscript{3} Eduard-Zintl-Institut f\"{u}r Anorganische und Physikalische Chemie, Peter-Gr\"{u}nberg-Str. 8,Technische Universit\"{a}t Darmstadt, Darmstadt 64287, Germany}

\begin{abstract}
Although the reverse non-equilibrium molecular dynamics (RNEMD) simulation method has been widely employed, the range of applicability is yet to be discussed. 
In this study, for the first time, we systematically examine the method against an unentangled melt of the Kremer-Grest type chain. 
The simulation results indicate that as the shear rate increases, the temperature and density become inhomogeneous. 
However, the average viscosity remains consistent with the results obtained using the SLLOD method under homogeneous temperature and density. 
We also confirm that the temperature-density inhomogeneity does not significantly affect polymer conformation.
\end{abstract}
\maketitle
\section{Introduction}
In molecular simulations, several approaches are used to predict shear viscosity. 
For zero-shear viscosity, the Green-Kubo formula is widely employed, wherein the viscosity is obtained from the velocity or stress autocorrelation function under equilibrium conditions\cite{D.todd}. 
In contrast, nonequilibrium simulations, which impose some external perturbation, are used for non-linear response under high shear, and the viscosity is directly obtained from the shear stress\cite{Lees}. 
In such simulations, the shear flow is maintained by a background flow field realized by a modified equation of motion\cite{D.todd,Bernardi}.

An alternative method proposed by M\"{u}ller-Plathe\cite{Florian,Florian_review} is the so-called reverse nonequilibrium molecular dynamics (RNEMD). 
This method imposes the momentum flux (or stress) and measures the gradient of the flow velocity. 
A significant advantage of the RNEMD method is that it does not require an external thermostat, as no external energy comes into the system. 
The system conserves momentum and energy.
Owing to this feature, the RNEMD method has been used in many systems, including monomeric\cite{Bordat,Guo,Soddemann,Manuella,Donko,Cavalcanti,Thomas} 
and polymeric\cite{Crusius,Eslami,Chen,XGuo, Oda,Nikoubashman,Chappa} systems.  
For instance, Chen et al.\cite{Chen} were the first to apply the RNEMD method to polymer melts.
They studied the viscosity of a coarse-grained polystyrene model and reported that the viscosity from the RNEMD method exhibits reasonable agreement with 
that from the SLLOD method from the Newtonian to higher shear rate regions. 
Nikoubashman and Howard\cite{Nikoubashman} extended the RNEMD method to a hybrid molecular dynamics method, where hydrodynamic interactions are explicitly considered. 
They calculated the shear rheology of semiflexible polymer solutions. 
Chappa et al.\cite{Chappa} applied the RNEMD method to their slip-spring simulations for entangled polymers.
Their results closely align with experimental data.  

Although previous studies have validated the RNEMD method, the range of applicability has yet to be discussed.
For instance, under the use of RNEMD, the temperature and density are not homogeneous but exhibit stationary quadratic profiles in the shear gradient direction. 
Consequently, the obtained viscosity is the average of the values under different temperatures and densities. 
This spatial inhomogeneity intensifies as the shear rate increases and potentially may cause problems according to the relation 
between the characteristic lengths of the inhomogeneity and the molecules. 
In an extreme case, when the dimensions of polymers are close to the size of the temperature-control domains where temperature is uniform, 
the dynamics of segments located at borders between different temperature domains may be disturbed.
If interactions between segments are relatively strong and the perturbation does not decay quickly, an unrealistic motion of polymers may appear.

This study examines the RNEMD method for a Kremer-Grest type polymer model\cite{Kremer}. 
For simplicity, we consider an unentangled melt. 
By systematically changing some control parameters of the RNEMD method, we discuss its applicability and accuracy in a wide range of shear rates by comparing with the SLLOD method\cite{D.todd,Bernardi}. 
We also compare the results from the $NVE$ and $NVT$ ensembles to see the effect of the thermostat. Details are shown below.

\section{Model and Simulation method}
We perform simulations for a Kremer-Grest type polymer melt. 
The polymer chains are composed of beads connected by finitely extensible nonlinear elastic (FENE) bonds. 
{All beads interact with each other via a truncated Lennard-Jones potential (LJ) or Weeks-Chandler-Andersen potential (WCA) that are purely repulsive. }
Truncated LJ and FENE interaction potentials are
\begin{align}
   U_{\rm{LJ}}(|\bm{r}|) &= \begin{cases}
     4\varepsilon\left[\left(\frac{{\sigma'}}{|\bm{r}|}\right)^{12} - \left(\frac{{\sigma'}}{|\bm{r}|}\right)^6\right] + U_c & \text{if } |\bm{r}| \leq r_c,\\
     0 & \text{if } |\bm{r}| > r_c,
   \end{cases}\\
   U_{\rm{FENE}}(|\bm{r}|) &= -\frac{1}{2}kR_0^2\ln\left[1-\left(\frac{|\bm{r}|}{R_0}\right)^2\right].
\end{align}
Here, $|\bm{r}|$ is the distance between the two beads, ${\sigma'}$ is the bead size, $\varepsilon$ is the intensity parameter, 
$k$ is the spring constant, and $R_0$ is the maximum extension length of the FENE bond. 
$U_c$ is a potential shift to ensure $U_{\rm{LJ}}(r_c) = 0$ at the cut-off length $r_c = 2^{1/6} {\sigma'}$, where the interaction becomes purely repulsive.
%The total potential energy of the chains is calculated by:
%\begin{equation}
% U_{\rm{total}}({\bm{r}_{i,k}})=\sum_{(i,k),(j,l),\text{pair}} U_{\rm{LJ}}(\bm{r}_{i,k} - \bm{r}_{j,l}) + \sum_{i=1}^{M}\sum_{k=1}^{N-1} U_{\rm{FENE}}(\bm{r}_{i,k+1} - \bm{r}_{i,k}),
%\end{equation}
%where $\bm{r}_{i,k}$ denotes the position of the $k$-th particle in the $i$-th chain. The first term on the right-hand side accounts for all pairs $(i,k)$ and $(j,l)$.

We choose length, energy, and mass units as ${\sigma'}=1$, $\varepsilon=1$, and $m=1$, respectively. 
We also set the Boltzmann constant, $k_B=1$, for convenience. 
In what follows, all physical quantities are expressed in these dimensionless units. 
The FENE potential parameters are the same as those in a standard Kremer-Grest type polymer model\cite{Kremer}: $k=30$ and $R_0=1.5$. 
The degree of polymerization (number of beads per chain) is $N=25$ {(This is below the polymerization degree between entanglements of the Kremer-Grest model of $N_e=30-90$\cite{pulz,uneyama,masubuchi}). }
The total number of polymer chains is $M=6400$. 
{The origin is located at the center of the simulation box. }
The simulation box sizes $L_x$ and $L_y$ are $91.8$ for the shear and shear gradient directions and $23.0$ for the vorticity direction. 
These dimensions are sufficiently larger than the equilibrium radius of gyration of a single chain, $R_{g,\rm{eq}} = 2.5$. The density of beads is $\rho = 0.826$.
 
\begin{figure}[H]
   \centering
   \includegraphics[width=100mm]{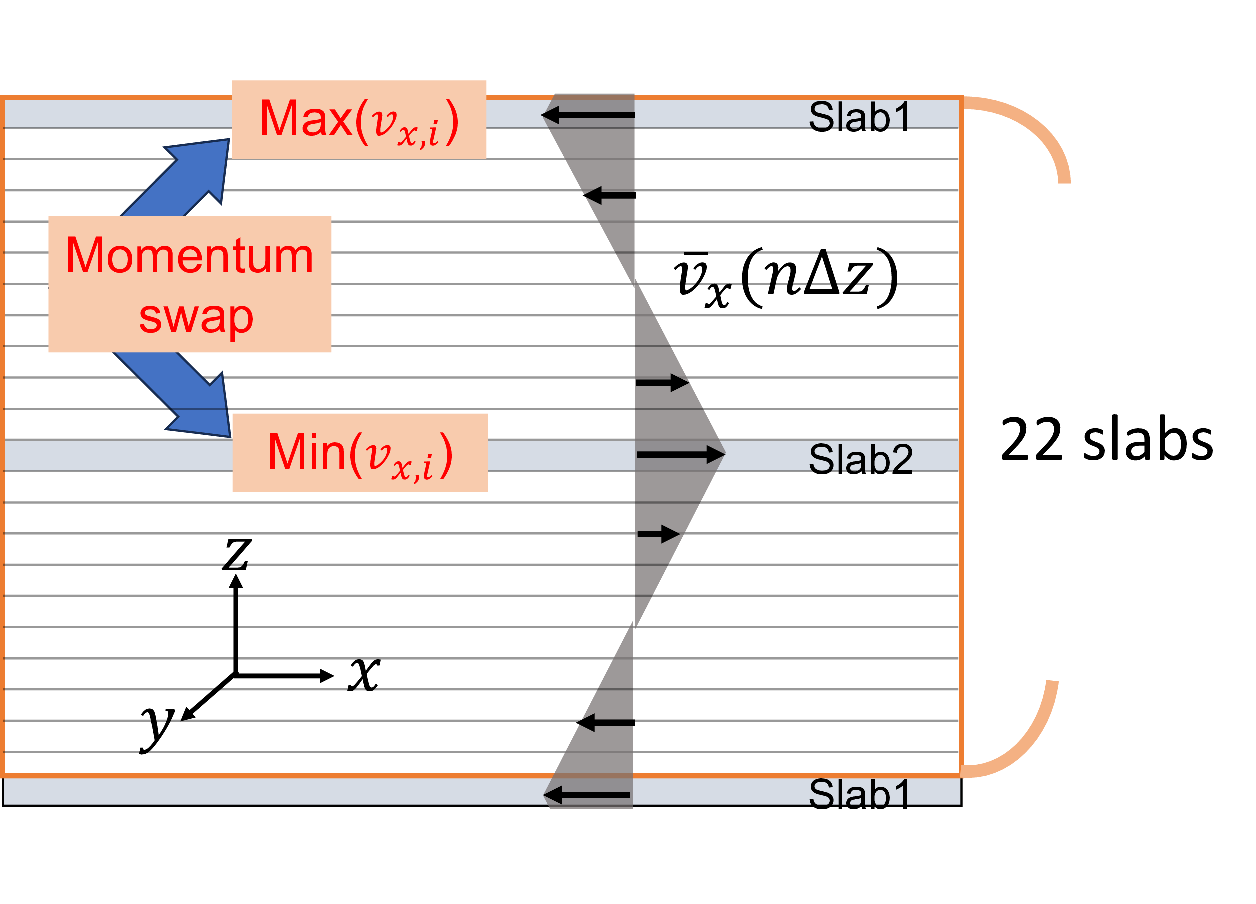}
   \caption{A schematic representation of the system and the momentum swapping in the RNEMD method. 
The orange frame indicates the simulation box, where the $x$, $y$, and $z$ axes are chosen towards the shear, vorticity, and shear gradient directions, respectively. 
The system is divided into 22 slabs in the shear gradient direction, and the upmost and middle ones are referred to as slab 1 and 2. 
Due to the periodic boundary condition, slab 1 is also shown at the bottom.  
The velocities of polymer beads are swapped between the bead that has the maximum positive velocity in slab 1 and the one with the maximum negative velocity in the shear direction in slab 2. }\label{image}
\end{figure}
%The flow field imposed on the system results in two symmetric planar Couette flows. 
%In this system, we set $3$-D periodic boundary conditions. 
After sufficient equilibration, we apply the RNEMD method to the system, which is periodic in all three dimensions, as illustrated in Fig.~\ref{image}. 
We impose the stress by swapping the velocities of beads. 
%Here, $x$-and $z$-directions are chosen along the shear and shear gradient directions. 
The velocity swapping is performed between beads that are selected in the following manner. 
We divide the simulation box into 22 slabs in the shear gradient direction, and we are supposed to realize two planar Couette flows, as shown in Fig.~\ref{image}. 
Specifically, in the $x$-direction, 
the slab at the top of the box (slab 1) is set to have the maximum negative flow velocity, and the one in the center of the box (slab 2) is chosen to have the maximum positive flow velocity.
{To realize such a velocity difference, 
we pick the particle with the largest positive velocity in the $x$-direction (the one most opposed to the intended flow) from the slab 1, and express its index as $i'$ and 
velocity as $v'$. }
In a similar way, we pick the particle with the smallest velocity in the $x$-direction from the slab 2, and
express its index and velocity as $i''$ and $v''$. 
{Then we swap the velocities of these two particles by setting $v_{i',x} = v''$ and $v_{i'',x} = v'$.}
This swapping is restricted to the exchange slabs and does not occur in the other slabs, which are used later to evaluate local velocity and temperature.
We control the frequency of this swapping to manage the resulting flow field, 
and the swapping frequency is defined by the number of swapping operations at one time $N_{\rm{s}}$ and the number of timesteps between swapping operations $W$.
For instance, if the value of $N_{\rm{s}}$ is $3$, we select $3$ bead pairs having the first, second, and third largest and smallest $x$ velocities at every $W$ time step. 
From $N_{\rm{s}}$ and $W$, we define a parameter $N_{\rm{r}}$ as
\begin{equation}
N_{\rm{r}}=\frac{N_{\rm{s}}}{W}.
\end{equation}
At each swap, we have a difference in the exchanged momentum for $k$-th steps of swapping operations and $l$-th number of swapping operations $\Delta p_{x,k,l}$:
\begin{equation}
\Delta p_{x,k,l} = m \left[ \max_{i \in (\rm{slab} 1)} v_{x,i} - \min_{i \in (\rm{slab} 2)} v_{x,i} \right].
\end{equation}
{Here, $m$ is the bead mass, which must be the same for each pair of particles. }
$\max_{i \in (\rm{slab} 1)}$ and $\min_{i \in (\rm{slab} 2)}$ represent the maximum and minimum velocities in the \rm{slab}s $1$ and $2$, respectively.
Accumulating $\Delta p_x$ for the entire simulation time, we have the total exchanged momentum rate $p_x$ written as
\begin{equation}
 \langle p_x\rangle=\frac{1}{t_{\rm{sim}}}\sum_{k = 0}^{t_{\rm{sim}} / \delta t}\sum_{l=1}^{N_{s}} \Delta p_{x,k,l}.
\end{equation}
Here, $\delta t$ is the integration step size and $t_{\rm{sim}}$ is the total simulation time.
With this $p_x$, we can evaluate the stress or momentum transfer rate $ \sigma$ as follows:
%In the RNEMD method, we can define the shear viscosity $\eta$ as the ratio of {steady-state stress $ \sigma$ and the velocity gradient $\partial \bar{v}_x/\partial z$ as follows:
\begin{equation}
  \sigma=\frac{\langle p_x\rangle}{2 L_xL_y}.
\end{equation}
%The statistical average $\left\langle ...\right\rangle$ is taken at the steady state.
{The factor 2 arises from the periodicity of the system, which allows the (conserved) momentum to flow from slab 2 to slab 1 through the liquid in the direction. }
From $ \sigma$, the viscosity $\eta$ is obtained as  
\begin{equation}
  \sigma=-\eta\dot{\gamma}\label{viscosity}.
\end{equation}
{Here, we assume that linear response is obeyed.}
Here, the shear rate $\dot{\gamma}=\partial v_{x}(z) / \partial z $ is calculated as fitting of $\bar{v}_{x,n}$:
\begin{equation}
 \bar{v}_{x,n}=\left\langle \frac{1}{\tilde{N}_n}\sum_{i=1}^{\tilde{N}_n}v_{x,i}\right\rangle ,
\end{equation}
where $\tilde{N}_n$, $v_{x,i}$, and  \(\bar{v}_{x,n}\) are the total number of beads, the component of the laboratory velocity of the $i$-th bead, and the average of the velocities of the beads 
in $(n-1/2)\Delta z \leq r_{z,i}< (n+1/2)\Delta z$.
%and $v_{x,i}$ is the component of the laboratory velocity of the $i$-th bead. 
%We calculate \(\bar{v}_{x,n}\) as the average of the velocities of the particles within the region \((n-1/2)\Delta z \leq r_{z,i} < (n+1/2)\Delta z\).
Here, $\Delta z$ is the thickness chosen as $\Delta z=0.5$, $n$ is an integer, and $r_{z,i}$ is the position of $i$-th bead in the shear gradient direction.
The statistical average $\left\langle ...\right\rangle$ is taken at the steady state.
{As shown later, the viscosity profile is nonlinear in the upper and central regions; thus, we evaluate the velocity gradient outside these regions; $-8.98 \leq z < -2.48$ (see Fig.~[7]) .}
Similarly, the local temperature and density $T_n$ and $\rho_n$ are defined for each slab as follows:
\begin{align}
T_n&=\left\langle\frac{1}{3\tilde{N}_n}\sum_{i=1}^{\tilde{N}_n}\left[\left(v_{x,i}- \bar{v}_{x,k}\right)^2+v_{y,i}^2+v_{z,i}^2\right]\right\rangle,\\
\rho_n&=\left\langle\frac{\tilde{N}_n}{L_xL_y\Delta z}\right\rangle, 
\end{align}
where $v_{y,i}$, and $v_{z,i}$ are the components of the laboratory velocity of the $i$-th bead. 
The resultant temperature and density profiles are parabolic, as shown later. 
When we perform $NVT$ simulations, the average temperature of the entire system is controlled with a Nos\'{e}-Hoover thermostat\cite{D.todd}.
Here, we set a temperature damping parameter to 2.0.

The equations of motion are integrated using the velocity Verlet algorithm\cite{Computer} in LAMMPS\cite{Lammps}. 
For equilibration, an $NVT$ simulation under quiescent state is performed for $1.0 \times 10^7$ steps with $\delta t = 0.005$ ($t_{\rm{sim}} = 5.0 \times 10^4$) at the temperature of 
$T=1.0$ controlled by the Nos\'{e}-Hoover thermostat\cite{D.todd}. 
Starting from the equilibrium state thus obtained, the RNEMD simulations are performed for up to $1.0 \times 10^8$ steps with integration step sizes of $\delta t = 0.005$ ($NVT$) and $\delta t = 0.0001$ ($NVE$), 
to realize the total simulation times of $t_{\rm{sim}} = 5.0 \times 10^5$ ($NVT$) and $t_{\rm{sim}} = 1.0 \times 10^4$ ($NVE$).
We note that the integration step size $\delta t$ in the $NVE$ ensemble is smaller than in the $NVT$ ensemble to stabilize the temperature.

For comparison, we conduct additional simulations using the SLLOD method\cite{D.todd} as implemented in LAMMPS to calculate the steady shear viscosity 
by imposing shear flow with Lees-Edwards boundary conditions\cite{D.todd}. 
The system is the same as that in RNEMD simulations. 
We use the Nos\'{e}-Hoover thermostat to control the temperature.
The integration step size is $\delta t = 0.001$, and the simulations are performed up to $1.0 \times 10^8$ steps ($t_{\rm{sim}} = 1.0 \times 10^5$).
 
\section{Result and Discussion}
\begin{figure}[H]
   \centering
   \includegraphics[width=85mm]{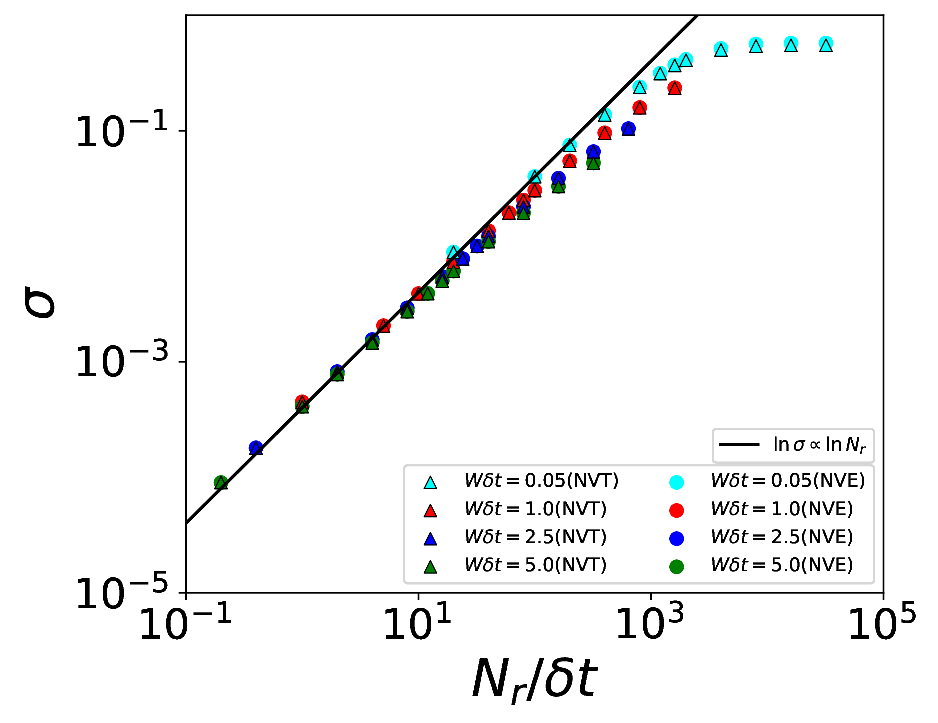}
  \caption{ Realized shear stress as a function of swap frequency parameter $N_{\rm{r}}/\delta t$ for $NVT$ and $NVE$ simulations with various swap intervals $W\delta t$. 
Black solid lines indicate $\sigma\propto$  ($N_{\rm{r}}/\delta t$) to guide the eye. The statistical error is smaller than the symbols in the plot.}\label{stress}
\end{figure}

\begin{figure}[H]
   \centering
   \includegraphics[width=85mm]{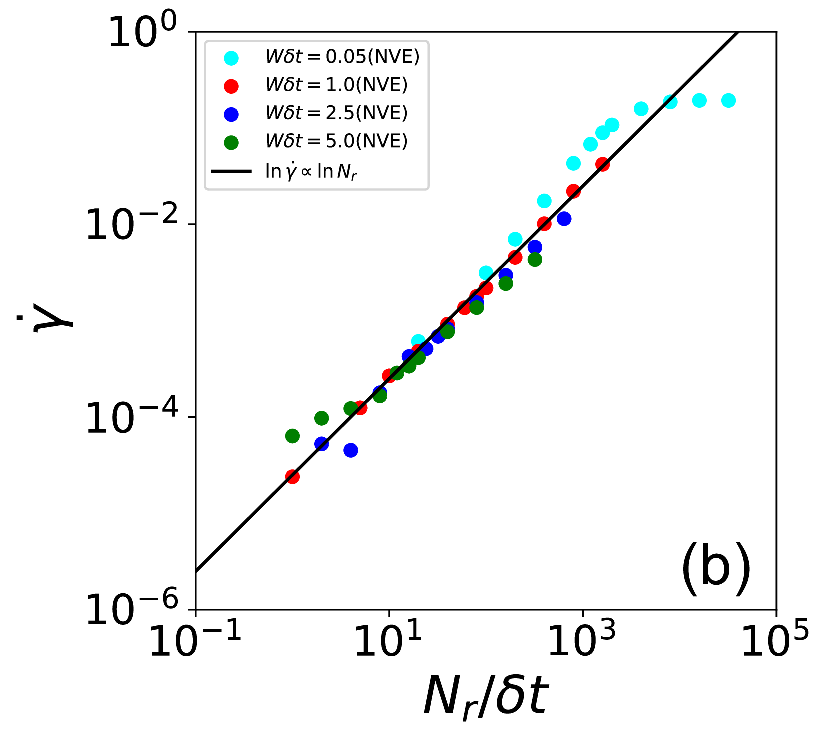}
  \caption{Realized shear rate as a function of swap frequency parameter $N_{\rm{r}}/\delta t$ for $NVT$ (a) and $NVE$ (b) simulations with various swap intervals $W\delta t$. 
Black solid lines indicate $\dot{\gamma}\propto$  ($N_{\rm{r}}/\delta t$) as a guide to the eye. }\label{shear_rate}
\end{figure}

\begin{figure}[H]
   \centering
   \includegraphics[width=85mm]{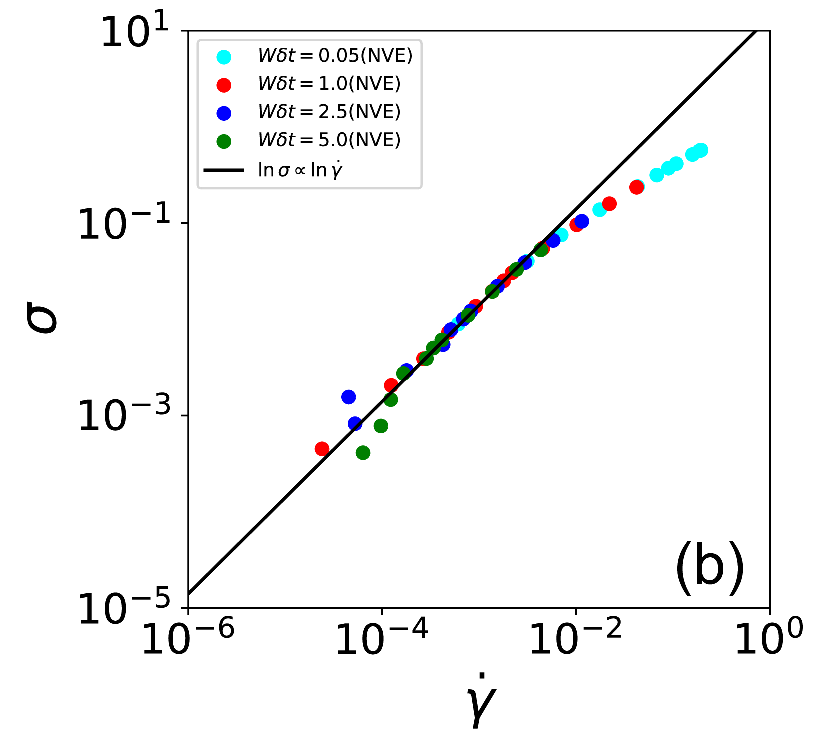}
  \caption{Stress versus shear rate for $NVT$ (a) and $NVE$ (b) simulations. Black solid lines indicate $\sigma\propto \dot{\gamma}$ for eye guide. }\label{shear_rate_stress}
\end{figure}

\begin{figure}[H]
   \centering
   \includegraphics[width=120mm]{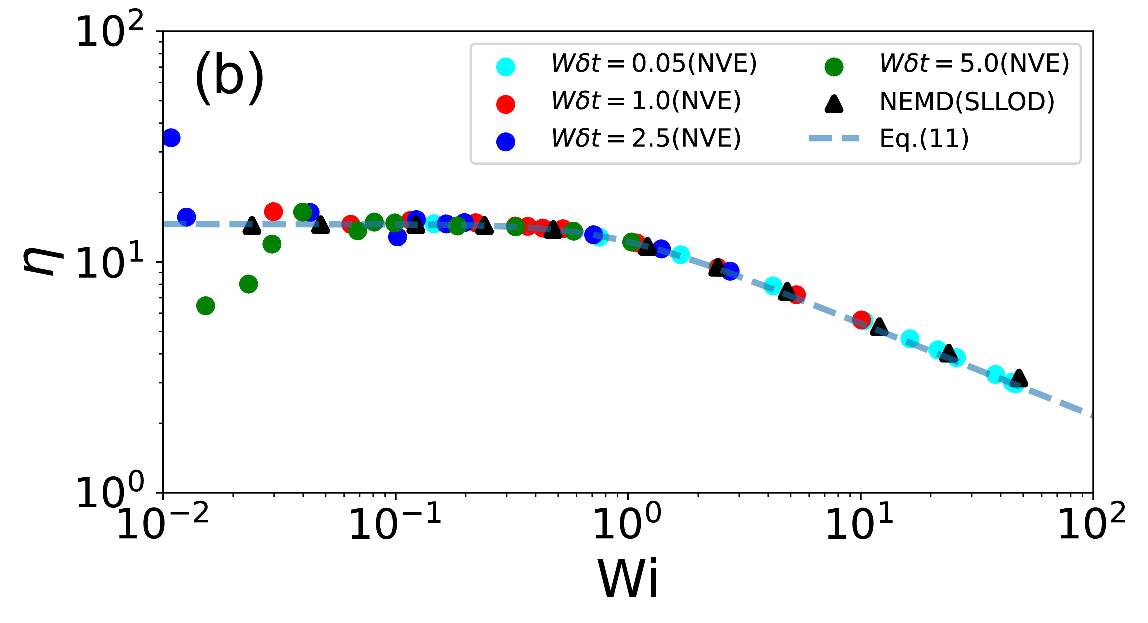}
   \caption{Viscosity plotted against Wi for the $NVT$ (a) and $NVE$ (b) ensembles, respectively. Circles represent the results from the RNEMD simulations with various $W\delta t$ values. Triangles are the results from the SLLOD simulations. Broken curves represent eq.~\eqref{power_law} with $\tau_c=2.85\times 10^2$ and $\alpha=0.20$.}\label{data_viscosity}
\end{figure}

In the RNEMD method, the shear rate and shear stress cannot be directly controlled, but they are realized via the momentum swap, as mentioned above. 
Figures~\ref{stress} and \ref{shear_rate} show the shear stress $\sigma$ and shear rate $\dot{\gamma}$ 
resulting from a given swap frequency parameter $N_r$ divided by $\delta t$ in $NVT$ and $NVE$ ensembles. 
The results demonstrate that stress and strain rate monotonically increase with the swap frequency. 
Namely, for small $N_r/\delta t$ ($\lesssim 50$), the stress is roughly proportional to $N_r/\delta t$, as shown by black solid lines. 
The shear rate is also proportional to $N_r/\delta t$. 
They deviate from a linear relationship for $\left( N_r/\delta t\right)>10^2$ and gradually saturate to plateau values.
However, this behavior cannot be solely described by $N_r/\delta t$, but also depends on the swapping interval $W\delta t$ in a complicated manner; with increasing $W\delta t$, 
the stress and strain rate slightly decrease. 
%Although this effect of $W\delta t$ is particularly observed for $NVT$ simulations, 
We can see there are no differences between the examined ensembles if the data are compared at the same set of ($N_{\rm{r}}/\delta t$, $W\delta t$) values. 
%c\begin{figure}[H]
%   \centering
%   \includegraphics[width=85mm]{./stress_NVT.eps}
%\end{figure}
\begin{figure}[H]
   \centering
   \includegraphics[width=85mm]{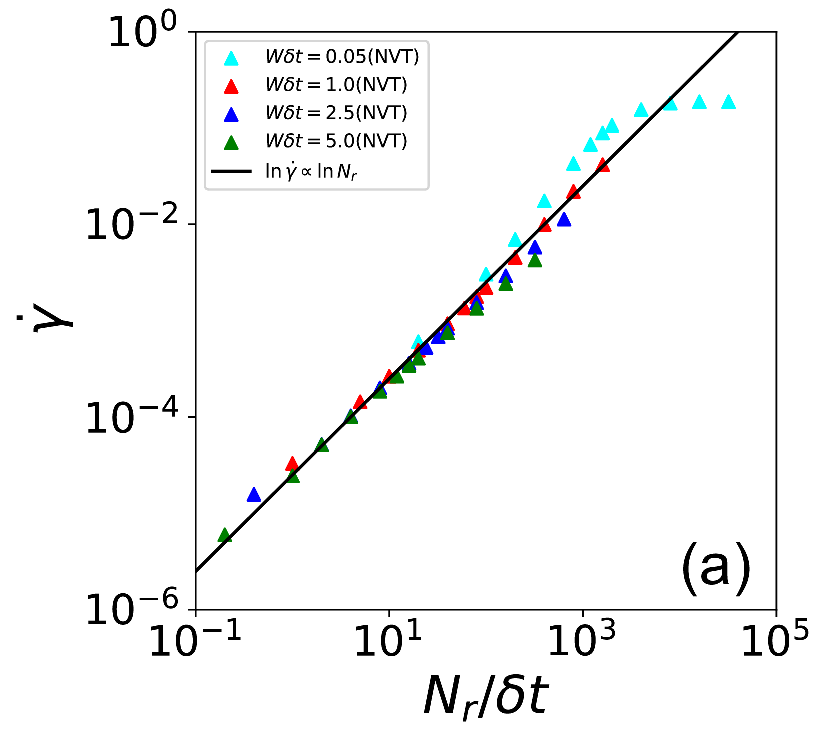}
\end{figure}
Nevertheless, all data points lie on a single master curve if they are plotted as the flow curve, where stress is plotted against strain rate. 
See Fig.~\ref{shear_rate_stress}. 
Even in the range of large $N_{\rm{r}}/\delta t$ values, for which the plots in Figs.~\ref{stress} and \ref{shear_rate} scatter depending on $W\delta t$, no deviation from a single curve is observed. 
{The relation is linear up to $\dot{\gamma}\approx 3\times 10^{-2}$.
We note that $NVT$ simulation seem to have a lower statistical uncertainty at very low shear ($\dot{\gamma}<10^{-4}$). }
\begin{figure}[H]
   \centering
   \includegraphics[width=85mm]{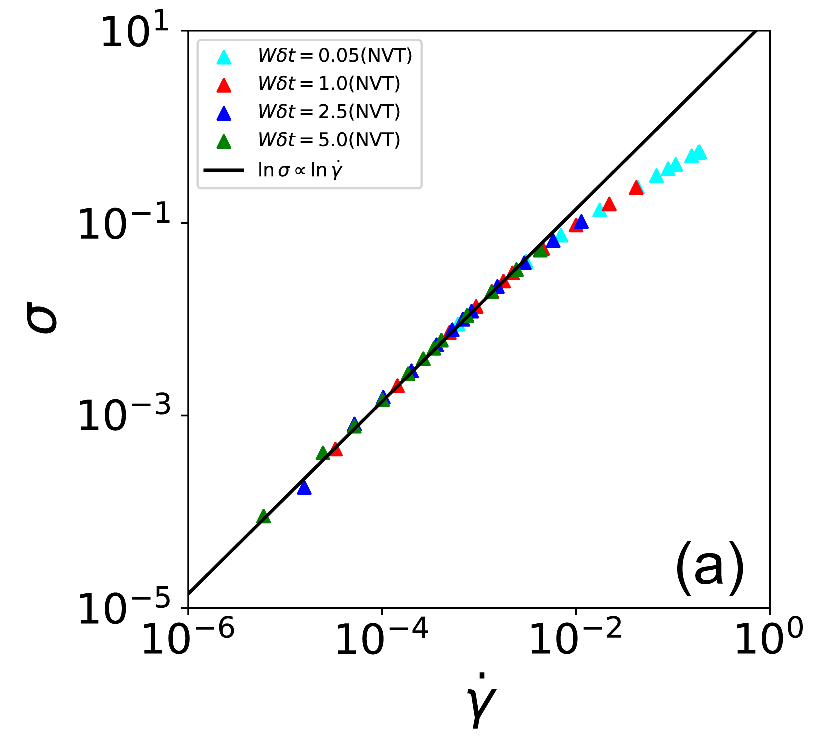}
\end{figure}
%From now, we discuss the Wi number dependence of $\eta$ over a wide range, and the characteristics of RNEMD itself by comparing with the SLLOD method.
Figure \ref{data_viscosity} shows the $\eta$ as a function of the Weissenberg number Wi in both $NVT$ and $NVE$ ensembles.
Note that the plotted values are according to the definition in Eq.\eqref{viscosity}, 
and they coincide with the values obtained from the flow curve shown in Fig.~\ref{shear_rate_stress} according to the relation $\eta=\sigma/\dot{\gamma}$. 
The Weissenberg number Wi employed here is Wi$=\dot{\gamma}\tau_R$, where $\tau_R$ is the viscoelastic Rouse time.
We determined the $\tau_R$ of the examined bead-spring polymer model at equilibrium as $\tau_R=2.4 \times 10^2$.
For comparison, the results from the simulations with the SLLOD method are also presented.
In Fig.\ref{data_viscosity}, all simulation results fall on a single master curve. 
When Wi is sufficiently small, the viscosity nearly matches the zero-shear viscosity, $\eta_0=1.49\times 10$, which was obtained from the linear relaxation modulus.  
As Wi increases, viscosity decreases, exhibiting shear thinning. 
\begin{figure}[H]
   \centering
   \includegraphics[width=120mm]{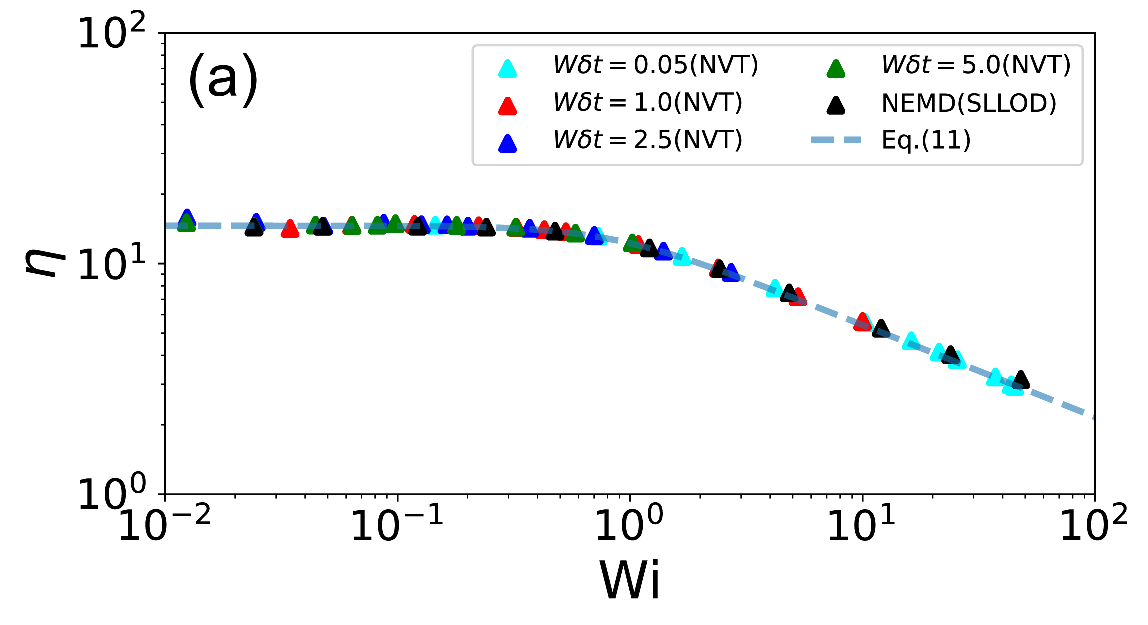}
\end{figure}
The obtained flow curve is similar to that found experimentally for unentangled polymers. Indeed, the broken curves in Fig.\ref{data_viscosity} indicate the empirical relation\cite{colby} shown below. 
\begin{equation}
 \eta(\dot{\gamma})=\frac{\eta_0}{\left[1+\left(\tau_c \dot{\gamma}\right)^2\right]^{\alpha}},\label{power_law}
\end{equation}
We determined the fitting parameters $\tau_c = 2.85 \times 10^2$ and $\alpha = 0.20$.
This indicates that in regions of high Wi, the viscosity follows a power law, $\eta \propto \mathrm{Wi}^{-0.40}$.
{We note that the cross-over time $\tau_c$ of the fit is comparable to the independently calculated Rouse time $\tau_R=2.4 \times 10^2$.}
{To validate our results, we also plotted the experimental data on unentangled polystyrene melts from Stratton\cite{Stratton} }
and the simulation results from Kr\"{o}ger and Hess\cite{martin} on a Kremer-Grest model for various numbers of beads per chain. 
Although our simulations slightly differ from previous research, there are some deviations within previous research itself, our simulations show reasonable agreement with these previous studies.

%This result differs from Colby et al, which states $\eta \propto \mathrm{Wi}^{-0.50}$ \cite{colby}.
%This result differs from the empirical relation proposed by Colby et al, which states $\eta \propto \mathrm{Wi}^{-0.5}$ \cite{colby}. 
%The discrepancy arises because our data includes higher Wi regions compared to Colby et al's data.
%To validate our results, we also plotted the results from Kr\"{o}ger and Hess\cite{martin} for different numbers of beads per chain. 
%They use the same polymer model as in our simulations. 
%Our results almost coincide with those of Kr\"{o}ger and Hess. when $N = 30$, indicating the reasonableness of our results. 
%However, it is evident that the power law depends on $N$. 
%Thus, the empirical relation proposed by Colby et al. corresponds to the case where $N \approx 60$ in our model.
\begin{figure}[H]
   \centering
   \includegraphics[width=120mm]{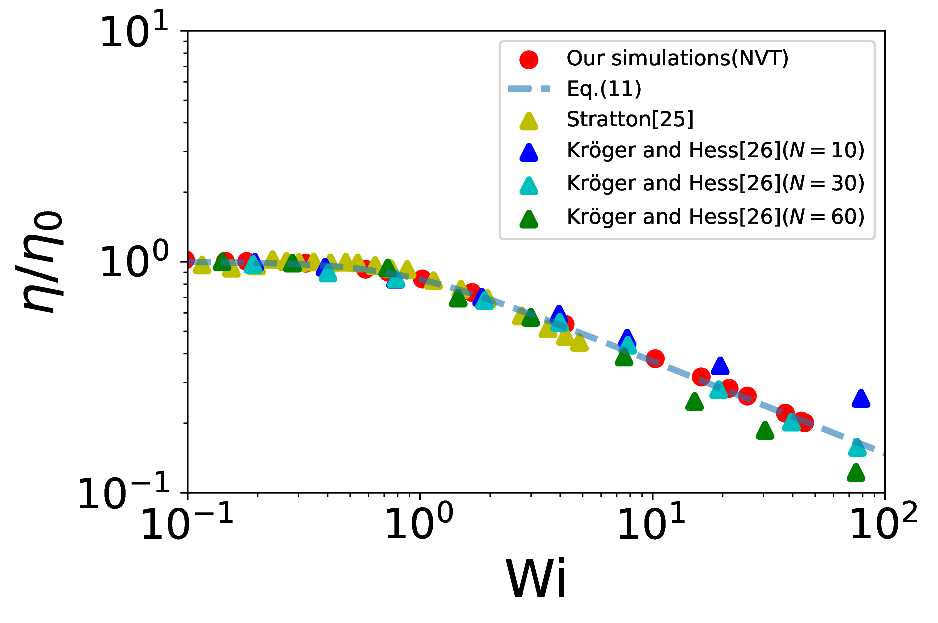}
   \caption{Comparison of our simulations of $NVT$ ensembles with previous research. Circles represent the results from the RNEMD simulations. Triangles are the results from the previous research. 
Broken curves represent Eq.~\eqref{power_law} with $\tau_c=2.85\times 10^2$ and $\alpha=0.20$. }\label{viscosity_compare}
\end{figure}

\begin{figure}[H]
   \centering
   \includegraphics[width=85mm]{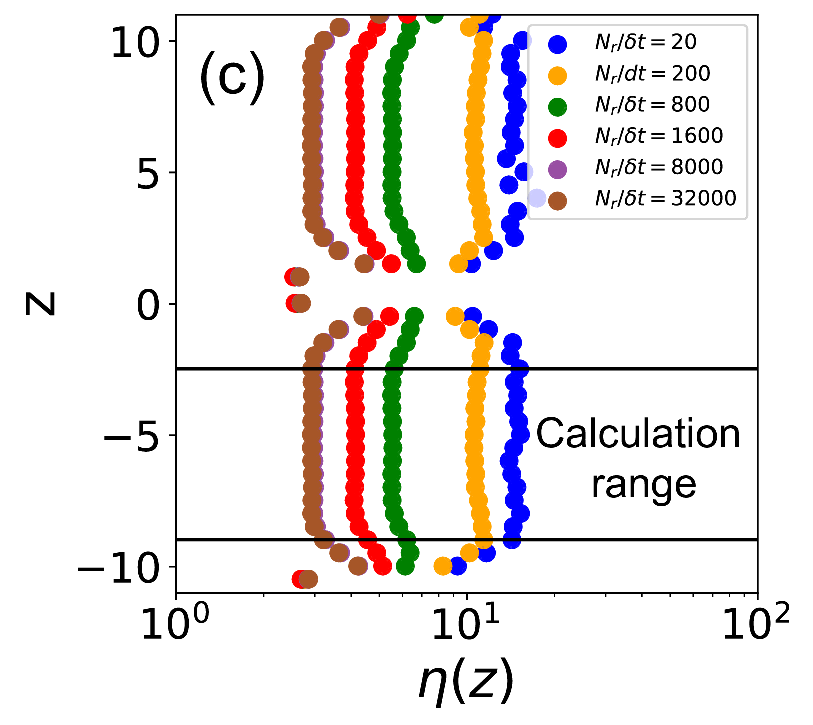}
\caption{Profiles of the temperature (a) density (b), and viscosity (c) along the shear gradient direction in $NVT$ ensembles at $W\delta t=0.05$. The magnitude of statistical error is smaller than the symbols in these plots.}
\label{Viscosity_profile}
\end{figure}

\begin{figure}[H]
   \centering
   \includegraphics[width=120mm]{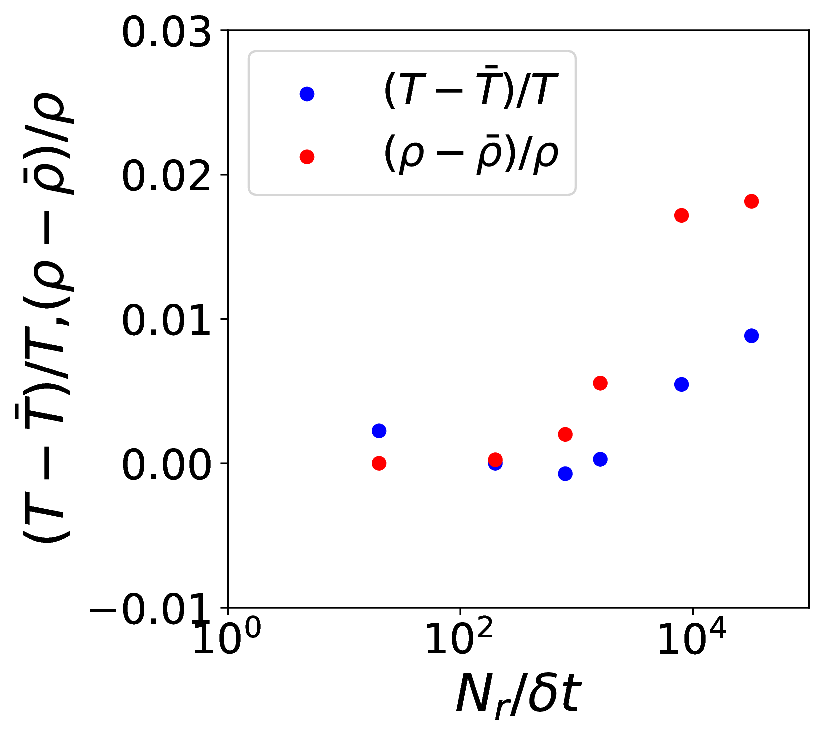}
   \caption{Deviation of the average temperature and density from target temperature and density in the calculation range for viscosity ($-8.98 \leq z < -2.48$) in $NVT$ ensembles at $W\delta t=0.05$. The magnitude of statistical error is smaller than the symbols in these plots.}
   \label{average_temp}
\end{figure}

\begin{figure}[H]
   \centering
   \includegraphics[width=85mm]{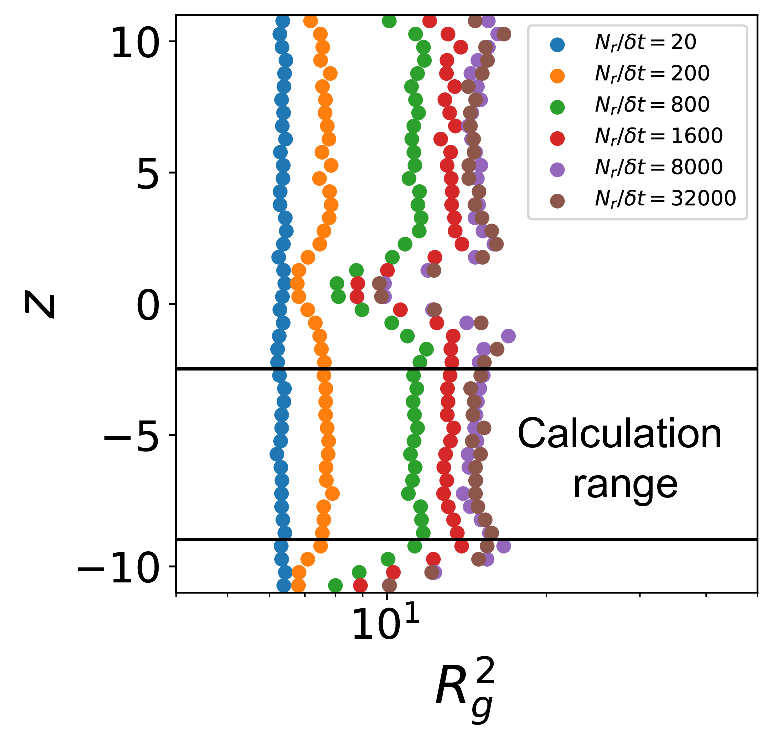}
   \caption{Profiles of squared gyration radius in $NVT$ ensembles at $W\delta t=0.05$. }\label{gyration_profile}
\end{figure}

\begin{figure}[H]
   \centering
   \includegraphics[width=85mm]{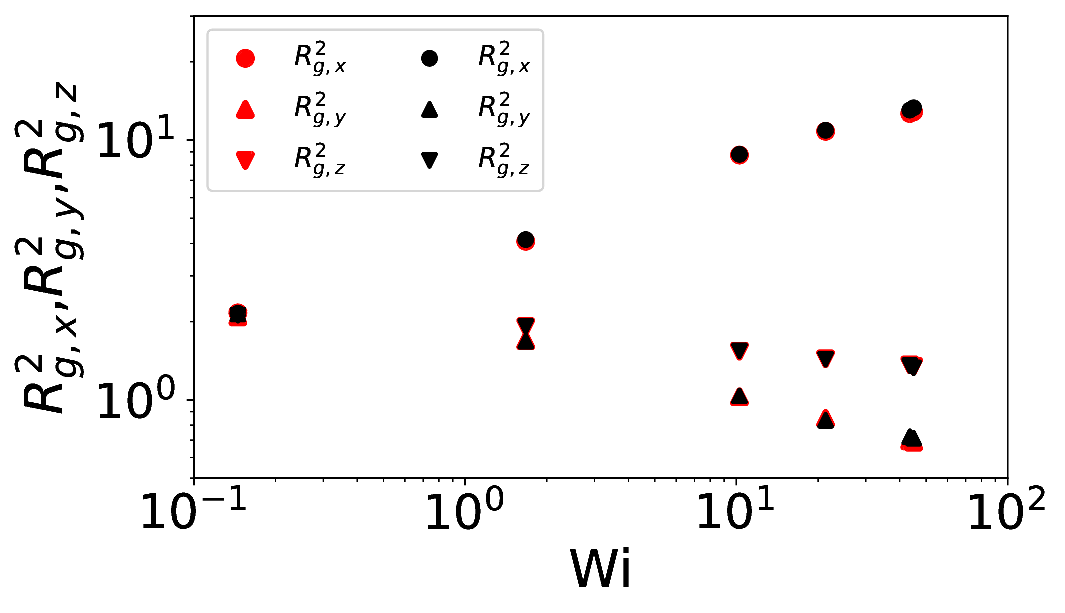}
   \caption{Cartesian components of the gyration tensor ($R_{g,x,n}^2 $, $R_{g,y,n}^2$, and $R_{g,z,n}^2$). The red color represents the results from RNEMD simulations in $NVT$ ensembles at $W\delta t=0.05$. The black color represents the results from SLLOD simulations. The magnitude of statistical error is smaller than the symbols in these plots.}\label{gyration}
\end{figure}

As mentioned in the introduction, the temperature, density, and viscosity are not uniform in the RNEMD method. 
Figure \ref{Viscosity_profile} exhibits these quantities averaged for each slab shown in Fig.\ref{image} in the steady state. 
We also show the calculation range for viscosity in $-8.98 \leq z < -2.48$. 
%(We do not include the upper range, such as $2.73<z<8.73$, because the results are not affected if we consider.) 
We only show the results in $NVT$ ensembles because the results are essentially insensitive to the difference of ensemble.
The temperature profile in Fig.~\ref{Viscosity_profile}(a) is parabolic in each Couette flow region, and the magnitude of inhomogeneity becomes more significant 
with a larger $N_r/\delta t$ value. 
In the central and edge slabs, there are artificial sharp spikes due to the velocity swapping. 
Following this temperature profile, the bead density is also parabolic, and it shows artificial peaks in the central and edge slabs where the temperature exhibits spikes. 
The viscosity profile shown in Fig.~\ref{Viscosity_profile} (c) is more complicated than that for the temperature and density. 
Here, we define the viscosity profile from the local velocity gradient.
There are sharp switchbacks around central and edge slabs, and the viscosity is almost flat in between these slabs. 
\begin{figure}[H]
   \centering
   \includegraphics[width=85mm]{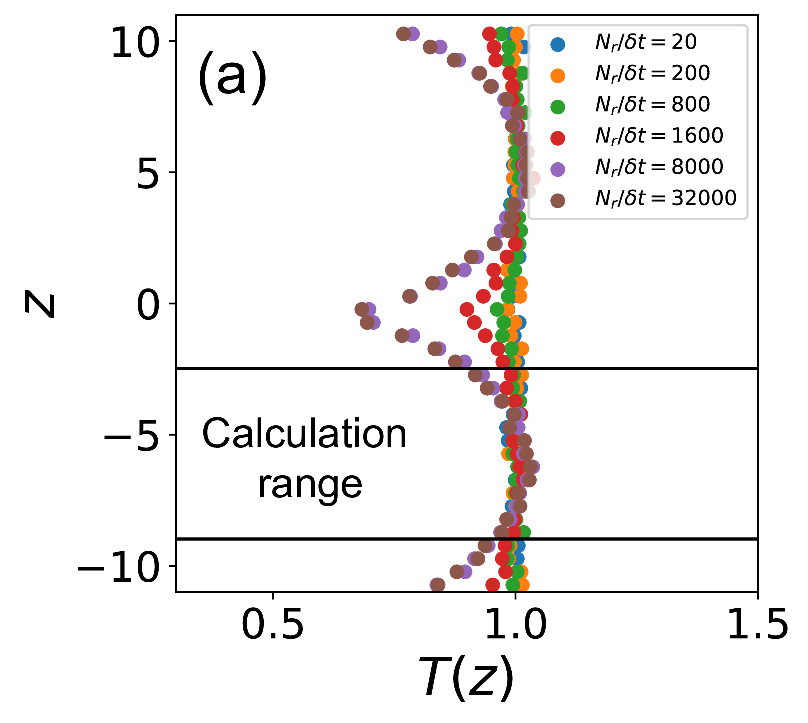}
\end{figure}
\begin{figure}[H]
   \centering
   \includegraphics[width=85mm]{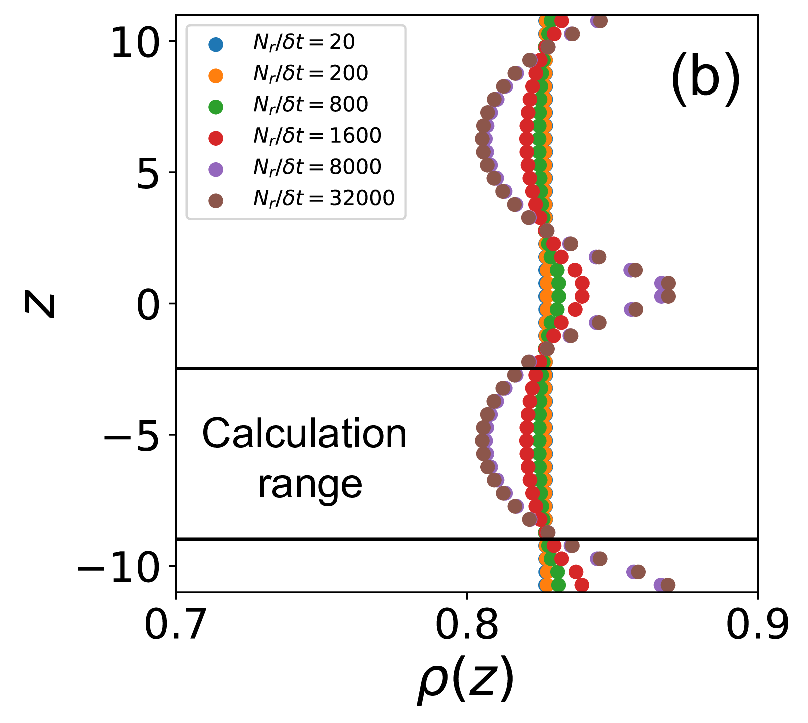}
\end{figure}
%From the profiles, we derive the average temperature and density in the calculation range because viscosity is a function of temperature and density. 
Figure \ref{average_temp} shows the relative deviation of the average temperature and density, $\bar{T}$ and $\bar{\rho}$,  from the target values, $T$ and $\rho$ in the region where the viscosity is calculated. 
%Here, we define the average temperature and density in the calculation range for viscosity ($-8.98 \leq z < -2.48$) as $\bar{T}$ and $\bar{\rho}$, respectively. 
%The equilibrium temperature and density are $T$ and $\rho$, respectively. 
We observe that both $(T - \bar{T})/T$ and $(\rho - \bar{\rho})/\rho$ increase as $N_r/\delta t$ increases.
This phenomenon is a characteristic of the RNEMD method due to the unphysical velocity swapping. 
To understand the effects of this phenomenon on the steady-state viscosity, we performed additional simulations with different densities\cite{Rouault}. 
(The deviation of the average temperature from target temperature is a minor effect compared with the deviation of the average density from target density.)
Although stress depends on the density, we find that the deviation from the average density in this study has a minor, negligible effect on the presented results.
This indicates that the decrease in $\bar{\rho}$ does not need to be considered when discussing steady-state shear viscosity under high shear rate regions.
{We also note that there are only small deviation at low swap rate ($N_r/\delta t<100$).
Systematic overestimation only sets in at high perturbation, where linear response is not valid either (see Figs.~[2-4]).}
We also observe the polymer conformation in terms of the gyration tensor defined as follows:
\begin{align}
 R_{g,\alpha,n}^2  &= \left\langle \frac{1}{N\tilde{M}_n} \sum_{i=1}^{\tilde{M}_n} \sum_{k=1}^{N}(r_{\alpha,n,i,k} - \bar{r}_{\alpha,n,i})^2 \right\rangle, \\            
 \bar{r}_{\alpha,n,i} &= \frac{1}{N} \sum_{k=1}^{N} r_{\alpha,n,i,k},         
\end{align}
{where $\tilde{M}_n$ is the total number of polymers in $(n-1/2)\Delta z \leq \bar{r}_{z,n,i} < (n+1/2)\Delta z$, and $r_{\alpha,n,i,k}$ is the coordinate of monomer $k$ in polymer molecule $i$ in the $\alpha$ directions.}
Here, the squared gyration radius is $R_{g,n}^2 = R_{g,x,n}^2 + R_{g,y,n}^2 + R_{g,z,n}^2$.
%The trace of the gyration tensor gives the squared gyration radius: $R_g^2 = \mathrm{Tr} \, \bm{S}$.
Figure~\ref{gyration_profile} shows the gyration radius profile in the steady state.
We see that $R_g^2$ remains almost constant in the calculation range for all RNEMD parameters.
This indicates that the inhomogeneous temperature and density fields do not cause inhomogeneity of polymer conformation in our simulations.
We also compare the polymer conformation with the SLLOD method.
Figure~\ref{gyration} shows the average diagonal gyration tensor elements in the calculation range for viscosity ($-8.98 \leq z < -2.48$) and the same for in the SLLOD calculations. 
The Cartesian components of the gyration tensor match with the SLLOD method across the whole range of the Wi. 
Consequently, we find that inhomogeneous temperature and density fields do not affect the polymer conformation.
{As a side result, we notice that upon shear, the polymer stretches in the direction of shear and contracts in the direction of the shear gradient and even more in the vortex direction. 
The overall expansion with increasing shear shows that the stretching in the shear direction outweighs the contraction in the direction perpendicular to it. 
This is in line with previous SLLOD simulations [\cite{kroger,,Xu,Gruel}] and RNEMD investigations of polystyrene melts of different chain lengths\cite{Chen}.}
%This is in line with previous RNEMD investigations of polystyrene melts of different chain lengths\cite{Chen}.}
\section{Conclusions}
We examine the reverse non-equilibrium molecular dynamics method for an unentangled melt of the Kremer-Grest type model under high shear. 
By increasing the frequency of the momentum swap, we achieve fast shear flows, with a shear rate of up to a Weissenberg number of approximately $100$. 
Although the temperature, density, and viscosity are inhomogeneous due to artificial momentum swapping, 
the average viscosity is fully consistent with that obtained from the SLLOD simulations under both $NVT$ and $NVE$ conditions. 
{The presence or absence of a thermostat does not significantly affect viscosity predictions in line with previous work on liquids\cite{Thomas}. }
We also investigated the inhomogeneity of temperature, density, steady shear viscosity, and gyration tensor. 
Our results suggest that deviations from the average steady-state viscosity and gyration tensor are minor on this scale. 
{The fact that the viscosity averages out to the correct value, despite the small temperature and density variations in the calculation regions, 
is in keeping with results obtained from RNEMD for molecular liquids\cite{Zhao,Xu,Gruel}.}
%This imply that it is unnecessary to consider the steady-state viscosity and gyration tensor profile within this scale.

%\bibliography{references}
%\bibliography{./../reference/water}
%\bibliography{./favorite}

%\bibliography{./../reference/water} 

\end{document}